\documentclass{ws-ijmpcs}

\begin{document}

\markboth{C.Bigongiari}
{Search for neutrino emission from gamma-ray sources with the ANTARES telescope}

%
\catchline{}{}{}{}{}
%

\title{SEARCH FOR NEUTRINO EMISSION FROM GAMMA-RAY SOURCES WITH THE ANTARES TELESCOPE}

\author{C.Bigongiari\footnote{
Instituto de F\'isica Corpuscolar, Apdo. de Correos 22085,
Valencia, E-46071, Spain.}}

\address{Instituto de F\'isica Corpuscolar, \\
Consejo Superior de Investigaci\'on Cientifica and Universitat de Val\`encia, \\
Edificios Investigaci\'on de Paterna,  Apdo. de Correos 22085, \\
Valencia, E-46071, Spain  \\
ciro.bigongiari@ific.uv.es}

%
%
%

\maketitle

\begin{history}
\received{Day Month Year}
\revised{Day Month Year}
\end{history}

\begin{abstract}
ANTARES is the first undersea neutrino detector ever built and
presently the neutrino telescope with the largest effective area
operating in the Northern Hemisphere.
A three-dimensional array of photomultiplier tubes detects the Cherenkov light 
induced by the muons produced in the interaction 
of high energy neutrinos with the matter surrounding the detector.  
The detection of astronomical neutrino sources is one of the main goals 
of ANTARES. 
The search for point-like neutrino sources with the ANTARES telescope 
is described and the preliminary results obtained with data collected 
from 2007 to 2010 are shown.  
No cosmic neutrino source has been observed and 
neutrino flux upper limits have been calculated for the most promising source candidates.

\keywords{High energy neutrinos; Cosmic rays; Undersea neutrino telescope. }

\end{abstract}

\ccode{PACS numbers: 95.85.Ry, 95.55.-n, 29.40.Ka}

\section{Introduction} 

Cosmic rays were firstly detected nearly one century ago and widely studied since then 
but their origin is still unknown.
High energy gamma rays have been detected from astronomical sources. In particular 
astronomical objects with relativistic outflows, such Blazars and Micro-quasars, 
have been firmly established as sources of high energy gamma rays~\cite{GammaRays}.  If these sources accelerate 
hadronic cosmic rays too they should produce also neutrinos which could travel unaffected to the Earth. 
The observation of point-like sources of cosmic neutrinos would 
provide a powerful insight into cosmic particle acceleration mechanisms. 

\section{Detection Principle}
 
The extremely low interaction cross-section of neutrinos with matter 
and the low neutrino fluxes expected from astronomical sources~\cite{NeutrinoFlux}
require large instrumented volumes to detect a statistically significant neutrino signal in a reasonable amount of time. 
Charged current interactions of high energy muon-neutrinos produce 
muons which can travel hundreds of meters before decaying. Muon neutrinos can therefore be 
detected even if they interact far away from the detector and their direction can  
be reconstructed with sub-degree accuracy because muon direction closely follows that of the parent neutrino.
The tracks of relativistic muons propagating in optically transparent dielectric media, like water or ice,
can be reconstructed detecting the induced Cherenkov light with a three dimensional array of photo-multipliers.  
The large background from down-going muons due to cosmic ray interactions in the atmosphere can be reduced 
by placing the detector deeply underwater/under-ice and selecting only up-going muons as neutrino candidates.
Up-going neutrinos produced in the interactions of cosmic rays in the atmosphere constitute an irreducible background.

\section{ANTARES Detector}

ANTARES is the first undersea neutrino detector ever built and
presently the neutrino telescope with the largest effective area
operating in the northern hemisphere~\cite{Antares}.
Located in the Mediterranean Sea 40 kilometers offshore the France coast near Toulon, 
the ANTARES experiment is predominantly sensitive
to neutrinos from the southern hemisphere in the TeV to PeV energy range.
The Cherenkov light is detected by an array of photomultiplier tubes, each housed in
a pressure resistant glass sphere~\cite{OpticalModule}. The photo-multipliers 
are grouped in triplets along strings which are anchored to the
sea bed at a depth of 2475 meters.  
ANTARES final configuration (may 2008) consists of twelve detection strings
spaced approximately 60 meters. 
Each detector string contains up to 25 triplets, separated by 14.5 meters.
The arrival time and charge of the photomultiplier signals are digitized~\cite{Ars} 
and transmitted to shore. Muon tracks are reconstructed by measuring the photon arrival times and 
the photomultiplier positions~\cite{Reconstruction}. 

\section{Data analysis}

The preliminary results presented here have been obtained analyzing data collected in 2007-2010 period with detector configurations containing between five and twelve detection strings. 
The overall live time is 813 days. 
Muon tracks have been reconstructed using a maximum likelihood fit of the observed photon arrival times. 
Tracks with bad reconstruction quality, quantified by a parameter based on the value of the likelihood function,
have been rejected. The angular uncertainty obtained from the fit has been also required to be smaller than one degree. 
Selecting up-going tracks only 3058 candidate neutrinos have been selected out of $\sim 100$ million reconstructed tracks. 
See \refcite{PointSources} for a detailed description of the analysis of data collected from 2007 to 2008.

\section{Search for point-like sources}

The search for point-like neutrino sources has been carried on 
looking for an excess of events over the atmospheric neutrino background anywhere in the field of view (all-sky search) 
and by testing the presence of a signal at the locations of 51 candidate sources (candidate list search). 
Candidate sources have been selected from catalogs of high energy gamma ray sources. 
The list includes also the most significant spot in the all-sky search performed by the IceCube collaboration
with the 22 string detector configuration. 
See table \ref{CandidateSourcesTab} for the list of candidate sources.

\begin{table}[ht]
\tbl{Candidate source list. The source coordinates and  
  the 90\%~C.L. limits on the flux intensity $\phi$ expressed in $10^{-8} \rm GeV^{-1} cm^{-2} s^{-1}$ units 
   are shown. 
}
{\begin{tabular}[ct]{l r@{.}l r@{.}l l | l r@{.}l r@{.}l l}
\hline
         Source
         & \multicolumn{2}{c}{$\alpha_s(^{\circ})$ }
         & \multicolumn{2}{c}{$\delta_s(^{\circ})$ }
         & \multicolumn{1}{c}{$\phi$ } 
         & \multicolumn{1}{c}{Source }
         & \multicolumn{2}{c}{$\alpha_s(^{\circ})$ }
         & \multicolumn{2}{c}{$\delta_s(^{\circ})$ }
         & \multicolumn{1}{c}{$\phi$ } \\
\hline
HESS J1023-575       &   155&83    &   -57&76    &  6.6  &   SS 433             &   -72&4     &     4&98    &  4.6 \\ 
3C 279               &  -165&95    &    -5&79    &  0.1  &   HESS J1614-518     &  -116&42    &   -51&82    &  2.0 \\ 
GX 339-4             &  -104&30    &   -48&79    &  5.8  &   RX J1713.7-3946    &  -101&75    &   -39&75    &  2.7 \\ 
Cir X-1              &  -129&83    &   -57&17    &  5.8  &   3C454.3            &   -16&50    &    16&15    &  5.5 \\ 
MGRO J1908+06        &   -73&1     &     6&27    &  0.1  &   W28                &   -89&57    &   -23&34    &  3.4 \\ 
ESO 139-G12          &   -95&59    &   -59&94    &  5.4  &   HESS J0632+057     &    98&24    &     5&81    &  4.6 \\ 
HESS J1356-645       &  -151&0     &   -64&50    &  5.1  &   PKS 2155-304       &   -30&28    &   -30&22    &  2.7 \\ 
PKS 0548-322         &    87&67    &   -32&27    &  7.1  &   HESS J1741-302     &   -94&75    &   -30&20    &  2.7 \\ 
HESS J1837-069       &   -80&59    &    -6&95    &  8.0  &   Centaurus\ A       &  -158&64    &   -43&2     &  2.1 \\ 
PKS 0454-234         &    74&27    &   -23&43    &  7.0  &   RX J0852.0-4622    &   133&0     &   -46&37    &  1.5 \\ 
IC22 hotspot  &    75&45    &   -18&15    &  7.0  &   1ES 1101-232       &   165&91    &   -23&49    &  2.8 \\ 
PKS 1454-354         &  -135&64    &   -35&67    &  5.0  &   Vela X             &   128&75    &   -45&60    &  1.5 \\ 
RGB J0152+017        &    28&17    &     1&79    &  6.3  &   W51C               &   -69&25    &    14&19    &  3.6 \\ 
Geminga              &    98&31    &    17&1     &  7.3  &   PKS 0426-380       &    67&17    &   -37&93    &  1.4 \\ 
PSR B1259-63         &  -164&30    &   -63&83    &  3.0  &   LS 5039            &   -83&44    &   -14&83    &  2.7 \\ 
PKS 2005-489         &   -57&63    &   -48&82    &  2.8  &   W44                &   -75&96    &     1&38    &  3.1 \\ 
HESS J1616-508       &  -116&3     &   -50&97    &  2.7  &   RCW 86             &  -139&32    &   -62&48    &  1.1 \\ 
HESS J1503-582       &  -133&54    &   -58&74    &  2.8  &   Crab               &    83&63    &    22&1     &  4.1 \\ 
HESS J1632-478       &  -111&96    &   -47&82    &  2.6  &   HESS J1507-622     &  -133&28    &   -62&34    &  1.1 \\ 
H 2356-309           &     0&22    &   -30&63    &  3.9  &   1ES 0347-121       &    57&35    &   -11&99    &  1.9 \\ 
MSH 15-52            &  -131&47    &   -59&16    &  2.6  &   VER J0648+152      &   102&20    &    15&27    &  2.8 \\ 
Galactic Center      &   -93&58    &   -29&1     &  3.8  &   PKS 0537-441       &    84&71    &   -44&8     &  1.3 \\ 
HESS J1303-631       &  -164&23    &   -63&20    &  2.4  &   HESS J1912+101     &   -71&79    &    10&15    &  2.5 \\ 
HESS J1834-087       &   -81&31    &    -8&76    &  4.3  &   PKS 0235+164       &    39&66    &    16&61    &  2.8 \\ 
PKS 1502+106         &  -133&90    &    10&52    &  5.2  &   IC443              &    94&21    &    22&51    &  2.8 \\ 
PKS 0727-11          &   112&58    &    11&70    &  1.9  &                      &             &             &      \\ 
\hline
\end{tabular}
\label{CandidateSourcesTab}}
\end{table}

\begin{figure}[ht]
\centerline{\psfig{file=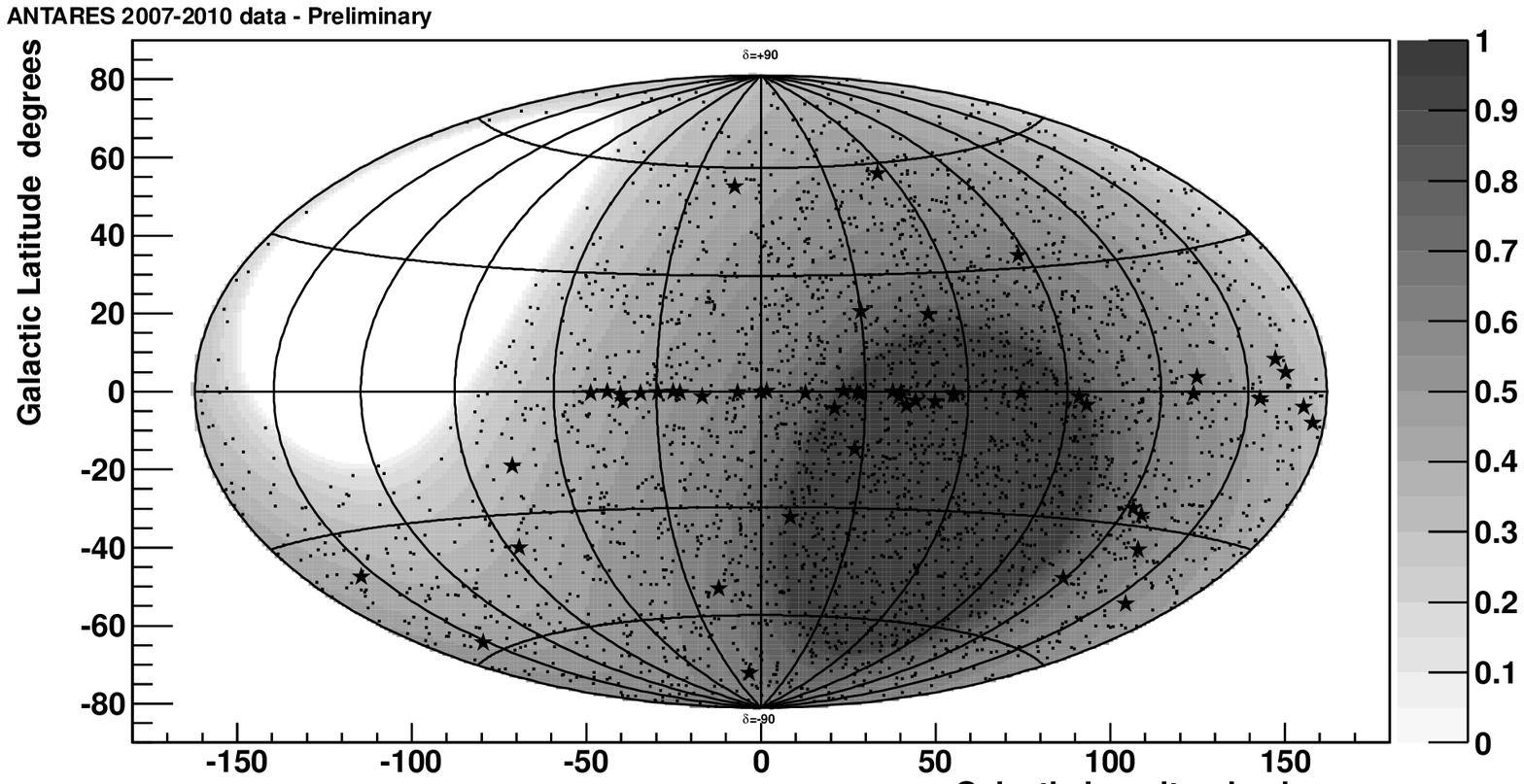,width=5.7cm} \psfig{file=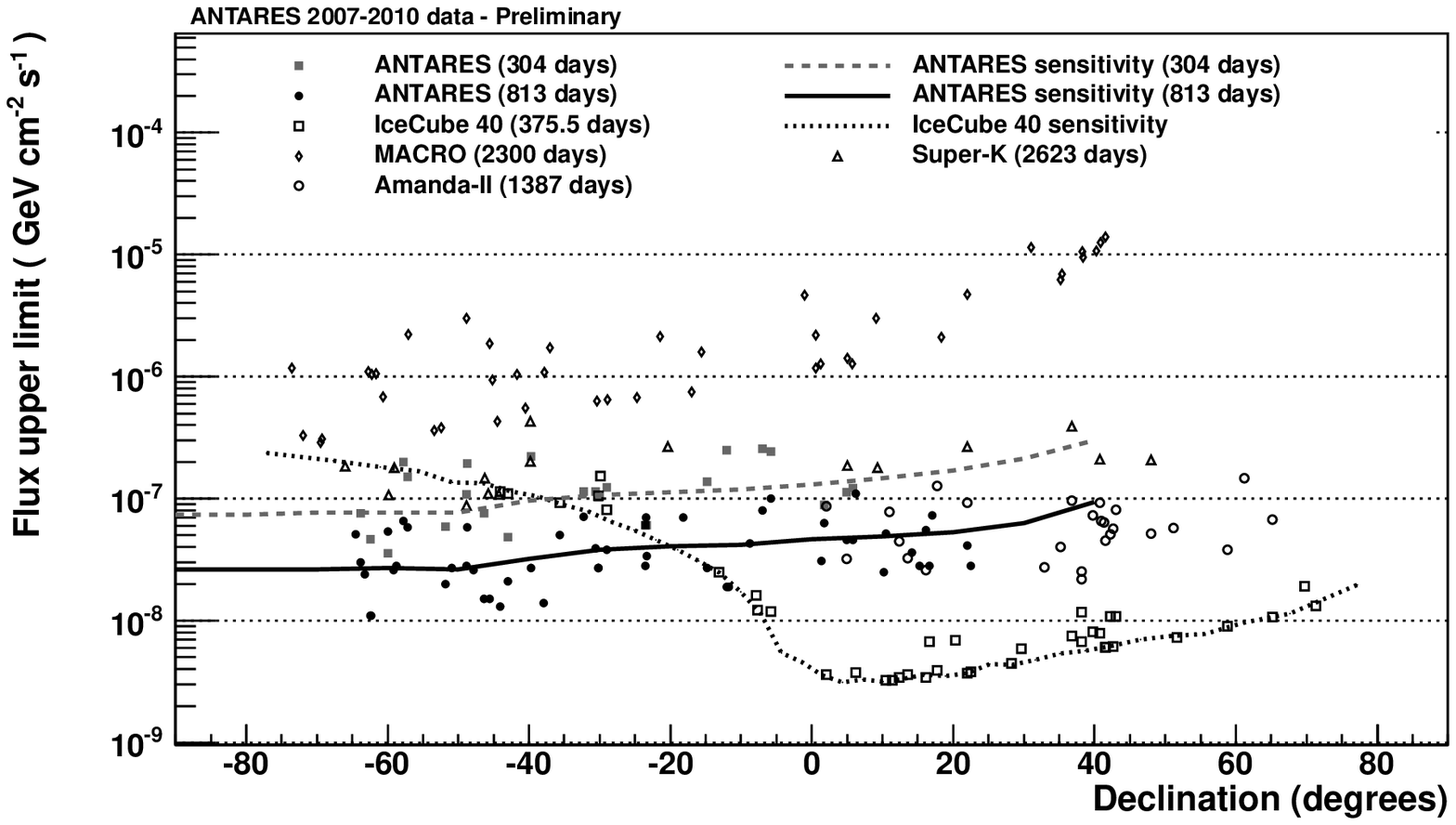,width=5.7cm}}
\vspace*{8pt}
\caption{Map in galactic coordinates of the 3058 selected neutrino candidates (dots) on the left. 
         The positions of the 50 source candidates are indicated by the stars.
         Neutrino flux upper limits for an $E_{\nu}^{-2}$ spectrum from selected candidate sources on the right. 
         The points show the 90\% C.L. limit at the declination of the candidate source. 
         Several previously published limits on source candidates are shown for comparison.
}
\label{SkyMap+Limits}
\end{figure}

No significant excess of events has been found at the location of any candidate source. 
The most promising source candidate is an unidentified gamma-ray source, HESS~J1023-575, while the second one is a well-known Blazar, 3C~279. 
Neutrino flux upper limits, calculated assuming a $E_{\nu}^{-2}$ spectrum, are shown in table \ref{CandidateSourcesTab}. 
Such limits are the most stringent to date for many candidate sources at $\delta_{S} < -20^{\circ}$.
No significant cluster of neutrino candidates has been found in the all-sky search either, see figure \ref{SkyMap+Limits}. 
The position of the most signal-like cluster is $\alpha_{S} = 46.49^{\circ} \; \delta_{S} = -64.97^{\circ}$ 
where five events are within one degree. Such amount of events is compatible with the background only hypothesis. 
The two-point auto-correlation function of neutrino candidates has been calculated  looking for an excess at any level of angular separation. 
No significant excess of events has been found at any angular scale. 

%
%
%

\section{Conclusions}

No significant excess of events has been found either with 
the full sky search or the candidate list search. Limits on the high energy
neutrino flux have been calculated for a number of selected source candidates. 
The limits obtained for many of the candidate sources at declination below $-20^{\circ}$
are the most stringent to date.

\section*{Acknowledgments}

The author gratefully acknowledges the financial support of the Spanish Ministerio de Ciencia e Innovaci\'on (MICINN), 
grants FPA2009-13983-C02-01, ACI2009-1020 and Consolider MultiDarkCSD2009-00064 and of the Generalitat Valenciana, 
Prometeo/2009/026.

\end{document}